\documentclass[11pt]{article}
\usepackage{setspace}
\usepackage{amssymb}
\usepackage{amsmath}
\usepackage{amsfonts}
\usepackage{slashed}

\def\be{\begin{equation}}

\def\ee{\end{equation}}

\def\dd{\partial}

\def\bea{\begin{eqnarray}}

\def\eea{\end{eqnarray}}

\newcommand\eps{\epsilon}

\makeatletter
\def\blfootnote{\xdef\@thefnmark{}\@footnotetext}
\makeatother

\begin{document}

\singlespace

\begin{flushright} BRX TH-6318 \\
CALT-TH 2017-18
\end{flushright}

\vspace*{.3in}

\begin{center}

{\Large\bf A brief history (and geography) of Supergravity: the first $3$ weeks...and after}

{\large S.\ Deser}

{\it 
Walter Burke Institute for Theoretical Physics, \\
California Institute of Technology, Pasadena, CA 91125; \\
Physics Department,  Brandeis University, Waltham, MA 02454 \\
{\tt deser@brandeis.edu}
}
\end{center}

\begin{abstract}
I summarize, at its $41^{\hbox{\tiny{st}}}$ -- and what would have been Bruno Zumino's $94^{\hbox{\tiny{th}}}$ -- birthday, the history of the discoveries of Supergravity, and some of its structure and later developments.
\end{abstract}

\section{Background}

Briefly, the relevant background -- and the ideas in the air -- prior to Supergravity's (SUGRA's) inception lay in two new, if quite different, realms: supersymmetry (SUSY), and the, also emerging, difficulties in achieving consistent gravity-higher ($s >1$) spin gauge field interaction. Indeed, the Western discoverers of SUSY, Julius Wess and Bruno Zumino [1], would frequently visit Boston from NYU to spread the SUSY gospel, which did get even our blas\'e attention after a while, especially since the simplest SUSY multiplet pattern $(s, s+1/2)$ linking adjoining Fermi-Bose fields had no obvious reason to stop at the then studied $s=0$ and $s=1/2$ models.  [The earlier, Eastern, discovery is [18].] The specific idea of a SUGRA was also in the air, both in terms of spacetime-dependent SUSY parameters [2] and of (albeit too general at the time) superspace approaches [2]. Separately, I had been intrigued for some time by the woes of higher spin gauge fields interacting -- as they must -- with gravity [2,3], and was particularly struck by Buchdahl's [4] early, if classical, study of the first difficult, $s=3/2$, case.

A rather different, but extremely potent, motivation was the striking improvement observed in the UV behavior of SUSY models, where infinities from the Bose/Fermi components miraculously cancel each other. It had been shown a bit earlier that quantum gravity coupled to matter was in dire need of miraculous cancellations, being non-renormalizable already at $1$-loop level for spin ($0,1/2,1$ -- both Maxwell and YM) matter sources [5], while the source-free system was known to diverge at the next, $2$-loop, level [6]. That UV hope was an enticing -- if still unresolved -- carrot; indeed, SUGRA not only shares the one-loop finiteness of pure GR [7] -- the only ``matter" field to do so, but stays finite at $2$ loops because there are no ($\sim$Riemann$^3$) invariant counterterms there. The (Riemann$^4$) stick, found a bit later [8], only strikes from $3$ loops on.
On a personal level, I had invited Zumino to lecture at the last -- 1970 -- Brandeis Theory Summer school. We stayed in touch thereafter, and his presence was a strong motivation to spend my Sabbatical term at CERN. After our initial SUGRA  work, we wrote a number of related papers in the field, including one on the 2D basis of ``Superstring SUGRA" [9].

\section{Action}
When I arrived on the evening of April 1,1976, Bruno was awaiting me in the CERN cafeteria and we instantly began our nonstop $3$-week endeavor -- especially at night, when we filled the CERN lecture hall's immense blackboards with our seemingly endless Fierz identity slogs! [We were once startled by another insomniac, Claud Lovelace, lurking in the curtains.] I had to go back to the US at dawn on the $22^{\hbox{\tiny{nd}}}$ for a short stay; fortunately, we were done with all calculations and writeup by then, and Bruno handed our manuscript [10] to Raoul Gatto, the local Editor of Phys. Lett. B, CERN's then ``house organ". One reason our work succeeded so rapidly was that we were both familiar with two papers by the great mathematician Hermann Weyl on coupling (Dirac) spinors to gravity [11] [Amusingly, this too was a West-East discovery; see [12]]. The first and oldest of these, showing how to do it, is well-known; the second -- directly relevant for us -- some two decades later, in 1950, was, and remains, an obscure gloss on the first. That was the year I began my student subscription to the Physical Review and pretended to read the fancy theory papers, in particular that one. Weyl noticed, in that short afterthought, that spinors can couple to affinities two ways, namely ``first" and ``second" order: either the affinities are regarded as independent variables or expressed as vierbein affinities, and that those two ways differed by (non-derivative) terms quartic in the spinors -- the torsion. Weyl's notation was rather cumbersome, and ours were real Majorana vector-, rather than Weyl's complex Dirac-, spinors, so we worked that dictionary out for ourselves, even though our aim was precisely to avoid dealing with the horrors of torsion, by sticking to first order. I was of course a rabid first order fan, as that was the basis of the 1959-62 ``ADM" formulation of GR, and Bruno was ambidextrous as well. [Indeed, we could have saved a lot of chalk by not worrying about the transformation properties of the independent (in first order) affinity, which is essentially just a Lagrange multiplier.] The beauty of our approach (pace Boltzmann's dictum that elegance is for tailors, not physicists) is that the entire action in [10] is the two-term sum of the (first order) Einstein and (the modern version of) minimally gravitationally coupled Rarita-Schwinger-Davydov-Ginzburg actions (yet another West-East discovery). The latter part just involves a covariant curl of the vector-spinor using the independent, non-metric, affinity, as does the Einstein (``Palatini") action. That's it -- neither quartic spinor nor auxiliary field debris! 
This deceptively naive form of the coupled field equations obeyed the required ``Bianchi" identities -- that their divergences vanished, both for the gravitational and spin $3/2$ field equations. That of the latter is mandated by SUGRA's invariance under the local, fermionic gauge transformation generalizing that of the flat space massless $3/2$ field (again, simply replacing an ordinary partial- by a covariant- derivative)  while the vierbein transforms like all SUSY bosons, with no derivatives of the gauge parameter. The vanishing of the spin $3/2$ field equation's divergence is in fact related to the other influence I mentioned at the start, Buchdahl's [4] remark that the massless spin $3/2$ equation is only consistent in Ricci-flat spaces, because its divergence is proportional to the $2$-index Ricci tensor that arises from the resulting covariant derivative commutator. He thought this meant this matter field could not consistently live in a General Relativity background (except as a test field), since the Ricci tensor was proportional to its stress tensor, so could not vanish. What this classical study omitted of course was the power of Fierz identities in (necessarily) quantized vector-spinor fields to show that this stress-tensor contribution vanishes identically (as it had better, not being gauge-invariant by itself). It also sealed the fate, as was later shown explicitly [15], of ``hypergravities", systems like $(2,5/2)$ because the spin $5/2$ tensor-spinor field equation's divergence has more indices, hence unavoidably brings in the full Riemann (rather than just Ricci) tensor, whose Weyl tensor part is left undetermined by the Einstein equations. This completes the (admittedly all too compressed, but accurate) survey of the motivation, history, and not least -- conceptual plus technical -- tools (and sweat, at our already advanced physics ages) involved in [10].  The appendices provide some essential details.    
 
\section{Re-action}
When we started, Bruno had just received (in those pre-internet days) a preprint of the initial version of the other SUGRA group [16], attempting to stake a claim to ownership of SUGRA, based on lowest order coupling in the traditional second order formalism. However, there is a long, instructive, history of would-be extended gauge invariant systems: Lowest order coupling always works, only to fail at next order -- the real test of consistency only occurs after the presumed first step's effect impinges on the next one; indeed, the $(2,5/2)$ model is of this type. Further, the $(3/2,2)$ idea was in any case already very much in play, as mentioned earlier. After we had finished, Bruno phoned me to report that a second version of [16] had just arrived, stating that an extensive computer calculation of their quartic (in the fermions) remainder terms had finally shown them all to vanish; we instantly, and unilaterally, decided to cite their results in our published version. As we have seen, both results were entirely on a par in timing, while all aspects of the model's constructions were mutually independent technically and conceptually. This was yet another example of simultaneous scientific discoveries, unsurprisingly when the right ideas are in the air. Most familiar in recent times are the three separate inventions of the ``Higgs" in the sixties, and most relevant to our theme, all the East-West groups mentioned earlier.

There is usually a reasonably fair consensus in such cases as to who did what when 
and deserves what share of the credit. To be sure, at the essential, scientific, level, 
each group is intrinsically rewarded just by seeing the dazzling new light for the first time, and the real truth is graven in the Platonic heavens. Here on earth, however, a Whig twist occasionally prevails, as it seems to have done here. This anniversary provides a chance to correct it, as Bruno often wished. In the Appendices, we provide some of the technical
details to help non-experts appreciate the subject. The third Appendix is devoted to a brief survey of some of the many subsequent developments.

\section*{Acknowledgements}
This work was supported by grants NSF PHY-1266107 and DOE\#desc0011632. I thank my moderating encouragers, Mike Duff and Mary K. Gaillard, my unique translator Joel Franklin, and Paul Townsend for useful correspondence.
A condensed version of this work was posted on CQG+.  This journal's founding editor, W. Bieglb\"ock (especially) and a referee both suggested useful improvements.

\section*{Appendix A - First versus second order formulations}
Before turning to SUGRA'€™s specific characteristics, we provide a reminder of the different expressions for a system's dynamics, a very ancient story. The original form of Lagrangian dynamics was in terms of a system'™s ``œposition"€ degrees of freedom, $q(t)$, with Lagrangian
\begin{equation}
L(q,\dot q) =T-V= 1/2 m \dot q^2 -V(q)     
\end{equation}
in the simplest case of second order time evolution equations, derived
as the stationary points of the action 
\begin{equation}
A= \int dt L ,       
\end{equation}
namely 
\begin{equation}
m \frac{d^2 q}{d t^2}  = -\frac{dV}{dq}.
\end{equation}
The next step is to reduce these equations to first order ones by introducing the momentum $p$ conjugate to $q$ through the new Lagrangian
\begin{equation}
L=p \dot q-H(p,q), H= p^2/(2m) +V(q) 
\end{equation}
whose resulting equations of motion are now doubled, but first order,
\begin{equation}
p=m \dot q, \, \, \, \, \, \, \, \, \, \, \,  \dot p = - \frac{dV}{dq} , 
\end{equation}
completely equivalent to (3). This elementary reminder generalizes to cover field-systems with an infinite number of excitations. [One can also cope with equations of higher time derivative order, by further enlarging the above $(p,q)$ phase space: the general procedure is due to Ostrogradsky [17], after whom the  extended variables are named. Fortunately, we will not need this procedure for our, purely second order, applications.] 
Consider the simplest gauge theory, electrodynamics, whose traditional action is 
\begin{equation}
A_{\hbox{\tiny Max}} [A_\mu] =  -\frac{1}{4} \int F_{\mu\nu} F^{\mu \nu}, \, \, \, \, \, \, \, \, \, \, \,   F_{\mu\nu}(A)= \dd_\mu Aˆ'_{\nu} - \dd_\nu A_{\mu} 
\end{equation}
with resulting second order field equations 
 \begin{equation}
\dd_{\nu} F^{\mu\nu}(A)=0. 
\end{equation}
Letting $F_{\mu\nu}$ be an independent variable in
\begin{equation}
A[F, A_\mu] = -\frac{1}{2} \int [F^{\mu\nu} F_{\mu\nu}(A)- \frac{1}{2} F^2]  
\end{equation}
produces the pair of first order equations
\begin{equation}
F_{\mu\nu} =F_{\mu\nu}(A),  \, \, \, \, \, \, \, \, \, \, \,  \dd_{\nu} F^{\mu\nu}=0    
\end{equation}
completely equivalent to  (6). In this case, and its nonabelian Yang-Mills extension, there is no difference in matter coupling, since that is
``minimal"€, to $A_{\mu}$, rather than to $F_{\mu\nu}$.

We are now ready for the Einstein-Hilbert General Relativity action in its vierbein, $e_{\mu a}$, incarnation. The Ricci tensor is defined via the spin connection $\omega_{\mu ab}$ where $a$, $b$ denote local, tangent frame, indices:
\begin{equation}
R_{\mu a} = e^{\nu b} R_{\mu\nu a b} = e^{\nu b} [ \dd_\mu \omega_{\nu a b} + \omega_{\mu a}^{\, \, \, \, \, \, c} \omega_{\nu c b} - (\mu \leftrightarrow \nu) ] 
\end{equation}
The standard second order forms of the Ricci and scalar curvatures
have $\Gamma$ and $\omega$ as derived quantities, namely as the metric Christoffel
connection $\Gamma^\alpha_{\, \, \mu\nu}(g)$ and vierbein affinity $\omega(e)$.

The scalar curvature $R$ is the trace of Ricci with $g^{\mu\nu}$ or $e^{\mu a}$ respectively. We use an ``€œindex-free"€ notation throughout in order not to obscure the essentials. In the second-order case, the field equations
from varying the actions 
\begin{equation}
A = \int \sqrt{-g} R(\Gamma(g)) =\int \hbox{det}(e) R(\omega(e)) \longrightarrow G_{\mu \nu}(g) = 0 = G_{\mu a}(e),
\end{equation}
the usual Einstein tensor vanishing. Note that unlike the Maxwell €"vector model, GR has not two, but three tiers, $(g, \Gamma, R)$ resp 
$(e, \omega, R)$. The first order formulations are quite similar to the Maxwell case, namely $\Gamma$ or $\omega$ are now independent variables, like $F$ and $A$ there. [It obviously makes no sense to treat the $R$ as independent!]  We are now to vary (11),  but with the affinities as independent, along with $g$ and $e$. Varying with respect to the latter yields the  ``€œEinstein"€ equations, $G_{\mu\nu}=0=G_{\mu a}$, while variation of the affinities yield the ``€œmetricity"€ conditions $\Gamma=\Gamma(g)$, $\omega=\omega(e)$, doubling the number of equations as the price of first order.  All this is 
really identical to the primitive $(p,q)$ case, but with lots of indices!

Now, however comes the big difference: matter coupling. Whereas particles or scalar and vector fields all couple minimally --- to the metric only --- €"spinors necessarily require covariant derivatives, €"i.e., affinities, 
$D_\mu \sim (\dd_{\mu}+\omega )$ acting on the spinorial, matter, variables. When Dirac introduced spinors in the late twenties, their coupling to gravity had to be given; Weyl (and Fock [12]) provided the full mathematical treatment in 1929 [11], using the metric/vierbein affinities, that became the standard. Then, some two decades later [11], Weyl returned to the question --- €"as mentioned in text --- with a short treatment using the independent affinities instead, and deduced that the two methods differed by matter terms bilinear in the spinors,  $\omega\sim \omega(e)+ \kappa^2 \bar \psi  \psi$ that introduced the notion of torsion. He further showed that, for spin $1/2$, the respective actions would differ from each other by, quite nasty-looking, terms quartic in $\psi$. To summarize, spinor fields coupled to gravity (as they must be) thereby acquire an ambiguity, by choosing either first or second order form in the gravitational variables, leading to a difference in the respective actions quartic in the spinors. What Weyl did not treat
was the fact that the quantum matter fields obeyed certain ``€œFierz"€ identities that are crucial to SUGRA'€™s existence, as we shall see below
and of course dealt with spinors, rather than its spin $3/2$ vector-spinors.
This Appendix provides the necessary background for the formulations of SUGRA in the sequel.

 \section*{ Appendix B Supergravity (SUGRA)}

 We provide here a brief introduction to SUGRA in its original versions; the many subsequent generalizations to higher dimension and internal symmetries can only be briefly noted (there exist many introductions to SUGRA, e.g.\ [13]). 
The unlikely origin of ``Super"€ models is a novel symmetry initially linking adjoining spin $(s, s+1/2)$ theories via a fermionic constant gauge parameter $\alpha$, discovered in [18] and later independently in [1], with no bosonic $(s, s+1)$ analog, incidentally. The particular such model of interest here links $(3/2, 2)$, later extended to the full range $(0, 1/2, 1,3/2 ,2)$. Because  constant spinor
makes no sense in curved space, SUGRA can only exist if the gauge parameter becomes a local one, $\alpha(x)$. The simplest such doublet is
clearly the non-interacting scalar $(0, 1/2)$, whose (flat space) action is the sum of the corresponding free actions,
\begin{equation}
A = -\frac{1}{2} \int \left[S_{, \mu}^2 + i \bar\psi \slashed\dd S\right] , 
\end{equation}
easily verified to be invariant under 
\begin{equation}
\Delta S \sim -\bar\alpha \psi, \, \, \, \, \, \, \, \, \, \, \, \Delta \psi \sim -i \slashed\partial S \alpha, 
\end{equation}
using $(\slashed\dd)^2 = \Box$. The length dimensions of $(\alpha, S, \psi)$ are respectively $(1/2, -1, -3/2)$. 
The well-known free (linearized) $s=2$ and $3/2$ actions follow the same pattern: 
\begin{equation}
A=\int R^Q -\frac{i}{2} \int \left[ \eps^{\mu\nu\sigma\tau} \psi_{\mu} \gamma_5 \gamma_\nu \dd_\sigma \psi_\tau\right]   
\end{equation}
where $R^Q$ is the quadratic part in $h_{\mu\nu}=g_{\mu\nu} - \eta_{\mu\nu}$ of the scalar curvature density, or of its vierbein form, $\psi_\mu$ is a vector-spinor and the gammas are the usual (constant) Dirac matrices. Note that each action is invariant under the separate LOCAL gauge transformations with the vector $X_\mu$ and spinor $\alpha$ functions
\begin{equation}
\Delta h_{\mu\nu} = \dd_{(\mu} X_{\nu)}(x), \, \, \, \, \, \, \, \, \, \, \, \Delta \psi_\mu = \dd_\mu \alpha(x)
\end{equation}
Consider now the fully nonlinear extension of the abelian action (14); the gravitational part is just the full $A=\int\sqrt{-g} R(\Gamma,g) =\int |e| R(\omega,e)$ with the, so far irrelevant, choice of first or second order affinities of Appendix A. The spinor action
is covariantized, and thereby coupled to gravity, by turning all $\eta_{\mu a}$ into $e_{\mu a}$ and most importantly, the $\dd$ˆ' in (14) into the covariant derivative, 
$D =\ddˆ' +\omega$. Note that the choice of $\omega$ rather than $\Gamma$ is necessary for spinors and that the vector index part of $\psi_\mu$ does NOT require a covariant differentiation, because it is a curl in (14), where ordinary and covariant derivatives coincide. So now the choice to be made is whether $\omega$ stays as $\omega(e)$ or becomes independent, namely Weyl's 1929 or 1950 choices. We [10] opted for the ``modern"€ version,
\begin{equation}
A= -\int \left[ |\det{e}| e^{\mu a} R_{\mu a}(\omega) - \frac{i}{2} \eps^{\mu\nu\sigma\tau} \bar\psi_\mu \gamma_5 \gamma^a e_{\nu a} D_\sigma(\omega) \psi_\tau\right].
\end{equation}
But is this a consistent model --- €"unlike most higher spins coupled to gravity?
One must check ``Bianchi" --- €"vanishing divergence --- €"identities for both systems. That for the Einstein equations is guaranteed by general covariance.

As a special case of Noether's theorem, local gauge invariance is equivalent to the existence of a corresponding ``Bianchi" identity:  If the variation of an action under a, possibly spinorial/tensorial function vanishes, then the corresponding ``gauge current" is conserved.  The usual Bianchi identity is simply the statement that if the coordinate variation of $A[e_{\mu\nu}, \psi_i] = 0$, where $\psi_i$ is a collection of ``matter" fields, i.e.\ that $A$ is a coordinate invariant, then the coefficient of its (vanishing) variation is the divergence of the full Einstein equation.  Now we extend this to the other local, spinorial $\alpha(x)$, invariance.  But that quantity is just the divergence of the spin $3/2$ field equation, 
\begin{equation}
D_\mu \eps^{\mu\nu\sigma\tau} \gamma_5 \gamma^a e_{\nu a} D_\sigma \psi_\tau=0.    
\end{equation}
In fact the easiest way to check that (17) holds is to vary (16), without bothering to vary $\omega$ since it is just a multiplier (see below).  There are just three contributions:  Varying the $e_{\mu\nu}$ in each term, and the $\psi_\mu$ in the spin $3/2$ part.  It is then a simple matter of $\gamma$-matrix gymnastics to verify that Ricci tensor terms, namely the Einstein $\alpha$-variation and the $[D_\mu, D_\sigma]$ from the $\psi$-variation cancel.  [Indeed, it was the loss of conservation of the pure $3/2$ system that led to Buchdahl's ``inconsistency" conclusion [4], since $R_{\mu\nu}$ is proportional to the spin $3/2$ stress tensor, a quantity that is not even $\psi$-gauge invariant.  That is the magic of Fierz!]. This leaves the $\alpha$-variation of the vierbein in the $\psi$-term, a cubic in $\psi_\mu$, whose vanishing requires a relatively simple Fierz identity application.  The $\alpha$-variation of the vierbein is the normal bosonic, ``algebraic" one $\Delta e_{\mu\alpha} = \kappa^2 \bar\alpha \gamma_\alpha \psi_\mu$.  The above proof is in fact perhaps the shortest possible.

The variation of the independent $\omega$ was worked out by us very laboriously; in fact it is not necessary to do any work at all, as $\omega$ is really just a Lagrange multiplier
whose variation tells us that $\omega =\omega(e) + \bar \psi \gamma \psi$, where the torsion
part involves various Dirac gammas. This means simply that $\omega$ varies like $\omega = \omega(e) + \hbox{``}(\hbox{torsion})\hbox{"}$ under this ``local super" parameter!  We have now completed the construction of a consistent SUGRA in first order form. To see how it would go in the standard $\omega = \omega(e)$ formulation, we note that the two actions differ by terms quartic in $\psi$. Thus one must show that the extra $\psi^4$ terms are
invariant in that approach; their variation involves $\sim \psi^5$ combinations. It was the latter that were shown, in [16], using the Fierz identities and a large computer program, to indeed vanish. Furthermore, there is no {\it a priori} ``natural" quartic unlike the natural first order quadratic form.  We re-note at this point why no higher
spin fermions such as spin $5/2$  $\psi_{\mu\nu}$, can lead to a consistent ``hypergravity" doublet: the $[D, D]$ commutator in the divergence of the tensor-spinor now allows for terms proportional to the full Riemann tensor, only the Ricci part of which can hope to vanish: the Weyl tensor contribution is completely arbitrary [15].

At the risk of repetition, we quickly summarize the conceptual and technical differences between first and second order formulations of SUGRA.  Most important, first order provides a natural ansatz for the covariant spin $3/2$ action, namely its minimally coupled form, purely quadratic in the $\psi_\mu$.  Technically, its verification avoids having to deal with the messy variations of the independent, Lagrange multiplier $\omega$, so the verification only requires some $\gamma$-matrix gymnastics and use of cubic Fierz identities in $\psi_\mu$.  Instead, second order form quickly reveals its minimal coupling is not sufficient.  This means having to introduce quartic terms algebraic in $\psi_\mu$, with no obvious hint of to which, if any (!), will work, given the now quintic Fierz identities that are required.  To be sure, the first order approach might have also required higher powers of $\psi_\mu$, but this absence is immediately checked.

To complete this survey of the simplest and first SUGRA, we treat now its cosmological constant extension, found shortly after in the first paper of [19]. The question then is how to introduce a term $\sim \Lambda \det{e}$; since $\Lambda$ is a new, dimensional, constant, it is not surprising that there must be a corresponding, dimensional, non-derivative term in the matter variables, which pretty uniquely looks like a spin $3/2$ mass term.  Specifically, the index structure turns out to be $\sim \sqrt{-\Lambda} \bar \psi_{\mu} \sigma^{\mu\nu} \psi_{\nu}$. What is unexpected is that the mass is proportional to $\sqrt{-\Lambda}$, a forced anti-deSitter sign with our conventions; unfortunately our universe seems to be a deSitter one. However, there is always a loophole --- as we already pointed out in [20] --- that allows cancelation, and indeed sign change to dS, of $\Lambda$, a topic of renewed current interest [14].  Introduction of $\Lambda$ also requires a minor and quite physical, extension of the Super transformation rules, and as we noted, despite the apparent spin $3/2$ ``mass" term --- indeed, because of it --- neither the graviton nor its partner are truly massive, but propagate on the (de Sitter) light cone [20].

\section*{Appendix C: The post-1976 flowering of SUGRA.}

It is totally impossible to summarize, let alone discuss, the $\sim 15,000$ papers
on SUGRA generated to date. Instead, I can only very briefly highlight some of the directions that have been explored, at that with many omissions of important work!  These include: 1.\ Different dimensions, 2.\ Higher internal symmetrics, 3.  Quantum effects: fighting the divergences, 4. Relation to superstring theory.
The original early papers are fortunately gathered in a $2$-volume annotated anthology [21] that provide an encyclopedic (in the authors' words) guide for the years 1976 to 1989; there also exist other compilations, but an updated list would be a welcome addition.

SUGRA models are highly dimension-dependent, because the number of
Bose and Fermi degrees of freedom (DoF) must match: in the original $D=4$ models, this was guaranteed because all massless spin $>0$ fields have two DoF there. Perhaps the first $D\ne 4$, and certainly the lowest dimensional, system was the, $D=1$, spinning particle [22], whose bosonic part $x^\mu(t)$ and fermionic $\psi(t)$ represents its spin. It is of course also the simplest, and is amusingly the (1st order) particle plus spinor action coupled to $D=1$ SUGRA, which has no dynamics, or even action, of its own but ensures that the particle's action is both coordinate- and local super-invariant. Its simplicity makes it is an excellent starting-point to experience SUGRA: the action is:
\begin{equation}
A = \int d\tau \left[ \pi \dot x - \frac{1}{2} e \pi^2 - \frac{i}{2} \psi \dot \psi - \frac{i}{2} \chi \psi x\right],
\end{equation}
where $e$ is the ``einbein" and $\chi$ its  SUGRA partner ``spin $3/2$" fermion, here enforcing the on-shell conditions $\pi^2=0= \dot \psi x$; we have omitted the vector world indices on $x^\mu$ and $\psi_\mu$, understood to be summed throughout.  Amusingly, there is again room for a ``cosmological" plus particle mass term, exactly as in $D=4$.

The next, $D=2$, case is of special interest because it describes the superstring in its intrinsic motion as a $D=2$ surface moving in spacetime [9].
The variables are the two-component $(i=t,x)$ position $A_i (t,x)$ and Majorana vector-spinor $\chi_i (t,x)$, whose action is required to obey on-shell conditions just like the $D=1$ case, and which will be enforced again by coupling to $D=2$ SUGRA, which again has no dynamics (recall that Gauss's theorem says the Riemann scalar density is a total divergence in $2D$!). We will not repeat the covariant action here, but note that it is a relatively straightforward extension of (19) to a $2D$ curved space and that it has the additional Weyl symmetry under conformal transformations beyond its coordinate and local
SUGRA invariance. Thus the superstring may be thought of as the matter field coupled to $D=2$ SUGRA; needless to say this formulation does not directly probe the deep facts that the space-time must have $D=10$ for quantum reasons.

Moving on to $D=3$, a dimension where SUGRA is still not dynamical because
the Ricci tensor is equivalent (double dual of) Riemann, so it has no freedom --- the Einstein equations imply spacetime is everywhere flat in absence of sources and completely determined by them otherwise. But there is a new and unexpected model here, topologically massive gravity (TMG) [23], which does have one massive DoF with quite unusual properties: it consists of the usual Einstein-Hilbert term plus the Chern-Simons (CS)  invariant, an integral of 3rd derivative (but harmless) order, the sum now being the sum of Einstein and CS, schematically  $A(\hbox{grav}) \sim \int [R(\omega, e) + \mu^{-1} \omega^3 ]$, where $\omega$ is again an independent field and $\mu$ has mass dimensions. The Fermionic companion is the field strength
$f^\mu =\eps^{\mu\nu\tau} (D_\nu \psi_{\tau})$ and the $s=3/2$ part of its action is [24]
 \begin{equation}
A(3/2) \sim \int [\bar\psi D(\omega) \psi + \mu^{-1} f \bar f ]
\end{equation}
suitably adorned with gamma matrices and dreibeine. The total action represents one DoF of each species, is parity-variant but SUGRA-invariant
(for $\omega$ independent!), When $\mu$ goes to infinity, one recovers the dull pure Einstein DoF-less $3D$ SUGRA.
We have already done $D=4$; the successive steps of higher $D$, while each somewhat different, reveal no qualitative change --- until, that is, one hits $D=11$. Recall that there are two requirements on a SUGRA theory: the Bose/Fermi excitations must balance and the spin cannot exceed $2$ because of the ``Buchdahl" problem that the ``Bianchi" identities on the higher spin equations cannot be maintained.  We will return to the balance question in the next section, but it suffices to note that $D=11$ is the highest $D$ that obeys these
constraints, as was shown by a group-theoretic counting in [25]; furthermore, $D=11$ SUGRA is unique and can be compactly written as the sum of Einstein,
spin $3/2$ and a new, $3$-form Bosonic field [26]; note that this is one dimension higher than superstrings live in. The new, $3$-form part, is relatively simple in terms of the field strength $F_{\mu\nu\iota\sigma}= \dd_{[\mu} A_{\nu \iota\sigma]}$, namely a ``Maxwell" term
$A \sim -\int F^2$ plus a CS-like $A= \int \mu^{-1} \eps^{1\ldots 11} F_{1..4} F_{5..8} A_{9.11}$, and their Fermion vector-spinor counterparts.
Needless to say, there is a whole industry devoted to dimensional reduction
and the relation of various rungs on the $D$-ladder. It is also perhaps deep that while Bosonic gravity exists in any $D$, SUGRA puts a relatively low ``cap" on it.
There is no mystery why: Nature (or at least theory) abhors spin $>2$.

Our next topic is that of extended SUGRAS, another chapter in the Bose/Fermi balance: consider lower spin Supersymmetric (SUSY) models, 
$(s, s+1/2)$ etc. When these are appropriately coupled to SUGRA, say in $D=4$,
the resulting now SUGRA theory now has more components, labelled by the
``internal" symmetries. In particular the widest range would include spins
$(0, 1/2, 1, 3/2, 2)$ with various multiplicities---ending with just one graviton, however. But there is a way to double the seeming ``$N=4$" structure. 
It essentially consists of ``running down the scale" again to make $N=8$ with a single maximal spin $2$; remember that in $D=4$, each $s>0$ (and Majorana spinor ) has $2$ modes. The totals are $(70, 56, 28, 8,1)$ fields of spin $(0, 1/2, 1, 3/2,2)$ with $128$ modes of each species (the $70$ scalars have only $70$ excitations). Note that unlike lower $N$ models, $N=8$ is unique [27] in that it cannot couple to lower spin SUSY matter; this is also true of the $D=11$ model, where there simply is no ``matter" but only the $3$ SUGRA fields. This of course also differs violently from bosonic gravity, that accepts all matter fields (of $s<2$). Higher $N$ SUGRA has also generated enormous work in various directions, that I cannot cover here, but can easily be traced via arXiv; we will, however, briefly discuss the quantum aspects of various SUGRAs.  The sheer algebraic complication of handling high $N$ systems has led to the use and elaboration of superfields that simultaneously represent many components, much like (hyper) complex numbers unify two variables. A dynamical extension is to study the possible gauging of these systems, for example the internal coupling of the various fields to the vector components, rather than simply leave them free.  Both of these are now major subfields.

Finally, we turn to the loop aspects of 
SUGRA.  It is---by construction---a quantum theory because the spinor
components must be quantized, hence also the Bosonic ones---since they rotate into each other. This means it has to face the problem of perturbative non-renormalizability of GR and its dimensional coupling---Newtonian/Planck---constant. But SUSY models have been spectacularly successful in the cancelation of infinities between Bose/Fermi modes, culminating in the, not only renormalizability, but finiteness, of  $N=4$ Super Yang-Mills theory [28], so there is still hope a priori! This has been a roller-coaster, still unfinished,  saga, as we shall see. First, a reminder of $D=4$ Bosonic GR's problems. On purely dimensional grounds, its one-loop counterterms are quadratic in curvature; however, as pointed out in [7], the Gauss-Bonnet identity expresses$\int \hbox{Riem}^2$ as the combination $\int [4 \hbox{Ricci}^2 - R^2]$. Hence these terms can all be removed, to the one-loop order at hand, by a field redefinition $\Delta g_{\mu\nu} = \kappa^2[a R_{\mu\nu}  +b R g_{\mu\nu}]$, whatever the specific coefficients, using the tree level action via  $\int \Delta g_{\mu\nu} G^{\mu\nu}$. However, this salvation becomes unavailable as soon as matter coupling, in [7] to scalar fields, is included. This in turn led to a search for the effects of the other then known matter systems, spinors, and vectors (but not yet vector-spinors!)---both Maxwell and Yang-Mills. Rather arduous calculations [5] showed that none of them, nor mixtures, produced infinities that were redefinable away. At two loops, GR was shown to become infinite, by heroic calculation, because those infinities explicitly included Riem$^3$ terms, that were not saved by Gauss-Bonnet [6]. It was here that the second miracle of SUGRA took place; the first was that at one loop it IS field-redefinably finite---basically, SUGRA is a single system like pure GR, so its infinities are also combinations of the field equations, using ``Super Gauss-Bonnet". The miraculous aspect of $2$ loops is that there is NO cubic super-invariant, unlike the quadratic one [8,29]. After our realization of this second miracle, we were tempted to use Fermi's dictum, that if something is true to first and second order it's always true, but fortunately held off before submitting. Instead, we remembered that there is an old four-index quantity called the Bel-Robinson tensor, $B_{\mu\nu\sigma\tau} \sim \hbox{Riem}^2$ that is conserved, and symmetric, and whose square $\sim \hbox{Riem}^4$ can be supersymmetrically extended to be a viable counterterm at $3$ loops [8]! (Basically, $B$ is a  tensorial version of the Einstein, and its SUSY extension is a version of the spin $3/2$, ``stress tensor"). Indeed, it has become the bane of all subsequent attempts at healing some version of SUGRA. But there is yet another, modern, twist: Thanks to great improvement in computational techniques, the maximal, $N=8$, system has been studied to many loop order---well beyond the $R^4$ $3$ loop hurdle---and remains finite so far! For the latest review as of now, see [30].
Whatever the outcome, is has been an exciting saga---it may stop at some
as yet unknown loop order, or the system may yet be finite, in which case
it will be even more mysterious: In the real world, SUGRA is ``maximally" broken, so what would be the moral of a finite result? There is clearly room for progress here!  Separately, there have been many intriguing connections between $D=10$ and $11$ SUGRA and $D=10$ string theory; I cannot expand on this vast subject here
except to note the interest of any deep connection between string and quantum field theories, especially since strings are finite while the 
divergences of SUGRA become, if anything, worse at higher $D$ because the
the loop integrals are higher power and the Planck constant is higher dimensional!

In conclusion, I quote Chou-En-Lai's reply to a query as to the effects of the then $200$ year-old French revolution: ``it's too early to tell": SUGRA is a beautiful set of theories, very much broken in the real world, yet with many useful and unexpected lessons in theory-building past and future.

\section{Important update, September 2019}

Quite recently, a private organization, the Fundamental Physics Prize Foundation, announced its  ``Breakthrough" award for SUGRA, but only to one group, with no mention whatever of Zumino's and my contribution. This blatant Soviet-style airbrushing out of history is particularly dishonest in view of: all the relevant available dated documents, Zumino's widow's (the distinguished theoretician Mary K. Gaillard) explicit eyewitness testimony, and the universal literature citations of both papers on a par (even in a major book by the award Committee's Chair and two of its members, and by the other group's authors). Absolutely no reason was (nor indeed could be) given for this blatant Breakdown.

\end{document}